\documentclass[12pt]{iopart}
\usepackage{graphicx}
\usepackage{amssymb}
\begin{document}
\title[The dual-dose imaging technique]{The dual-dose imaging technique: a way to enhance the dynamic range of X-ray detectors}

\author{Evangelos Matsinos and Wolfgang Kaissl}

\address{Varian Medical Systems Imaging Laboratory GmbH, T\"{a}fernstrasse 7, CH-5405 Baden-D\"{a}ttwil, Switzerland}
\ead{evangelos.matsinos@varian.com and wolfgang.kaissl@varian.com}

\begin{abstract}
We describe a method aiming at increasing the dynamic range of X-ray detectors. Two X-ray exposures of an object are 
acquired at different dose levels and constitute the only input data. The values of the parameters which are needed 
to process these images are determined from information contained in the images themselves; the values of two parameters 
are extracted from the input data. The two input images are finally merged in such a way as to create one image containing 
useful information in all its entirety. This selective use of parts of each image allows both the contour of the irradiated 
object to be visible and the high-attenuation areas to retain their image quality corresponding to the information contained 
in the high-dose image. The benefits of the method are demonstrated with an example involving a head phantom.\\
\end{abstract}
\pacs{87.57.Ce, 85.57.Nk, 87.59.Fm}
\noindent{\it Keywords\/}: cone-beam CT, flat-panel detector, image quality, dual dose, dual pulse\\
%\submitto{\PMB}
%\maketitle

\section{Introduction}

The high-quality volumetric reconstruction is the common aim in many of the modern imaging modalities. A number of issues are 
casually packed in the term `high quality', ranging from the suppression of the noise and of occasional artefacts to the 
enhancement of the contrast and spatial resolution. The dynamic range of the detectors, which are used in the data acquisition, is 
strongly linked to many of these issues.

Matsinos and Kaissl (2006) describe a method to calibrate one of the operation modes of flat-panel detectors (FDP) manufactured 
by Varian Medical Systems, Inc.~(VMS), Palo Alto, CA; that mode had been developed as a response to the need for increasing the 
dynamic range of the VMS X-ray detectors. The present paper introduces an alternative approach which is general and simple 
enough to implement in a variety of FDP types; a patent application, relating to this method, is currently under reviewing, 
see Dasani \etal (2006).

The dual-dose imaging technique may be outlined as follows. Two X-ray exposures of an object are obtained at different 
dose levels. Although this technique might also be used in case of scans in the future (via an implementation similar to 
the one proposed in Eberhard \etal (2005) in dual-energy imaging), these exposures are currently static (no gantry 
movement), the irradiated object occupying a fixed position in space. Digitised images, corresponding to the two 
exposures, are obtained with the use of an X-ray detector. The first dose level is selected in such a way as to yield 
a good-quality signal in the high-attenuation areas of the projected image (onto the detector). Due to limitations in 
the range of the output signal (dynamic range of the detector), it is inevitable that the pixel signal in this high-dose 
image will saturate in the low-attenuation areas of the projected image; in most cases, the low-attenuation areas correspond 
to the contour of the irradiated object. The second dose level is chosen in such a way as to avoid saturation within the 
projected image, keeping at the same time the dose as high as possible (to suppress noise). Finally, the two images are 
combined and create one image with good-quality information in its entirety. The low-noise high-dose pixel signal 
is used wherever reliable (not saturated); otherwise, it is substituted by the low-dose pixel signal after the latter has 
been scaled by a factor which is determined on the basis of the two input images.

Given the geometry of an imaging unit, the delivered dose is a function of the X-ray-tube (simply called `tube' hereupon) 
voltage, tube current and pulse width. A variation of the voltage in the two exposures induces physical effects which can 
only be accounted for by a dedicated calibration of the imaging unit, similar to the one developed by Matsinos and Kaissl 
(2006) in the case of the dual-gain mode; for instance, the scatter contribution and beam-hardening effects are energy-dependent. 
Although the technique of the dual-energy imaging was originally proposed (as a means to visualise simultaneously different 
parts of the anatomy) about twenty years ago, the research interest in this domain has hardly diminished; the corroboration of 
this interest is manifested by the amount of patent applications which are currently under reviewing, e.g., see Kump and 
Odogba (2003), Jabri \etal (2004) and Zhao \etal (2004). On the other hand, by varying either the current or the pulse width, 
one modifies the \emph{number} of incident photons, not their energy; due to this reason, the implementation of an intensity-variation 
scheme constitutes a less demanding problem. On the technical side, the solution which is assumed in the VMS imaging 
units pertains to the variation of the intensity of the incident beam via changes induced in the width of the X-ray pulses. 
This is the reason why the terms `short pulse', `long pulse' and `dual pulse' are often used in the VMS nomenclature as 
synonyms of `low dose', `high dose' and `dual dose', respectively; in the present paper, we will make use of the latter terms.

\section{Materials and methods}

\subsection{The imaging unit}

The data analysed here has been acquired at the VMS laboratory in Baden, Switzerland. The `On-Board Imager' system, 
comprising the imaging unit of a VMS machine which was recently constructed and put into operation to enable image-guided 
radiation therapy, has been used. The detailed description of this system may be obtained directly from the website of the 
manufacturer (`www.varian.com').

The X-ray source is the VMS model G242; it is a rotating-anode X-ray tube with maximal operation voltage of $150$ kV. The 
tube is driven and controlled by the X-ray generator. In the mode described herein, the generator has been programmed to 
trigger two X-ray pulses in succession, the short pulse preceding the long one.

The VMS PaxScan 4030CB amorphous-silicon FPD, which is currently used in the data acquisition, is a real-time digital X-ray 
imaging device comprising $2048 \times 1536$ square elements (pixels) and spanning an approximate area of $40 \times 30$ cm$^{2}$. 
In order to expedite the data transfer and processing, the so-called half-resolution ($2 \times 2$-binning) mode is normally 
used; thus, the detector is assumed to consist of $1024 \times 768$ (logical) pixels (pitch: $388$ $\mu$m). Due to the 
high sensitivity of the scintillating material (thallium-doped cesium iodide) and to sophisticated noise-reduction techniques, 
the low-dose imaging performance of this type of detector is remarkable, save for a small band ($2.91$ mm wide) neighbouring 
its borders (`inactive area of the detector'). The digitisation depth of this FPD type is $14$ bits. In reality, however, 
nonlinear effects are introduced at signals far below the $14$-bit limit. A thorough method to determine accurately the 
threshold value (the highest pixel signal which still fulfills the dose-signal linearity) for each pixel separately has been 
described in Matsinos and Kaissl (2006). In the present paper, one constant (same for all pixels) threshold will be extracted 
from the data; to a good approximation, this number is equal to the minimal of the values of the threshold map. In the data 
processing, if a pixel signal exceeds this constant threshold value, the pixel will be assumed saturated.

\subsection{The dual-dose imaging technique}

\subsubsection{Usefulness.}

High-quality images are important in a variety of applications. Restricting ourselves to medical imaging, an obvious 
requirement is that anatomic details in the region of interest be discernible as good as possible. As regions of interest 
frequently lie well within the irradiated objects, high-dose X-ray pulses are needed to probe their structure; otherwise, 
the resulting images look grainy and the details are masked by quantum and discretisation noise. Evidently, the delivery 
of high dose is needed in the high-attenuation areas of the irradiated object.

Unfortunately, the delivery of high dose solves one problem at the expense of creating another; the low-attenuation areas 
of the object will not be visible if the delivered dose exceeds the limit which is associated with signal saturation. 
To conclude, due to limitations in the range of the output signal, the good visualisation of one part of the irradiated 
object impaired the retrieval of information in another.

The dual-dose imaging technique solves the afore-described problem by combining parts of the two images in such a way 
as to create one image containing useful information in all its entirety. The selective use of parts of the two images 
allows both the contour of the irradiated object to be clearly visible (avoiding saturation effects) and the high-attenuation 
areas to retain their image quality corresponding to the high-dose image. Therefore, the application of the technique 
results in the increase of the dynamic range of the detector which is used in the data acquisition.

\subsubsection{Description of the algorithm.}

One of the main features of the present approach is that the combination of the two images is based on information which is 
exclusively contained within them; hence, no additional calibration is needed, that is, beyond the standard calibrations of 
the dark field (offset) and flood field (flat field, flatness or gain) which are invariably applied to all images. 

We now come to the description of the algorithm which is currently implemented. Two parameters are involved: hereupon, they 
will be called `threshold' and `ratio'; their values are obtained from the two input images. The threshold value determines 
whether the low- or the high-dose information is to be used in the combination of the two images; the ratio value determines 
the amount by which the low-dose image is to be scaled whenever it is used. In the combination of the two images, the norm 
is assumed to be the high-dose image. The threshold values obtained in the present paper are offset corrected; this is due 
to the fact that the input data have been corrected for offset and gain effects.

\begin{figure}
\begin{center}
\includegraphics [height=15.5cm,angle=270] {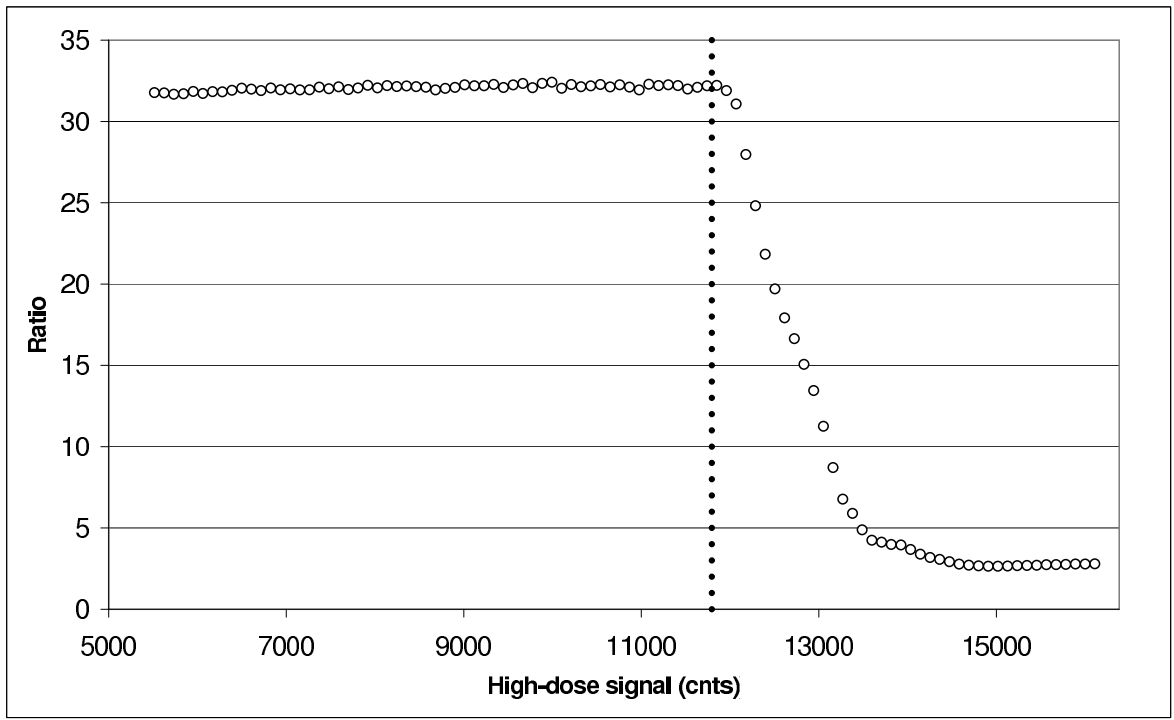}
%\vspace{-6cm}
\caption{\label{fig:RatioVSHighDose}The ratio of the pixel signals histogrammed in bins of the high-dose signal. The vertical 
dotted line corresponds to the threshold determined on the basis of an analysis of figure \ref{fig:Derivative}.}
\end{center}
\end{figure}

\begin{figure}
\begin{center}
\includegraphics [height=15.5cm,angle=270] {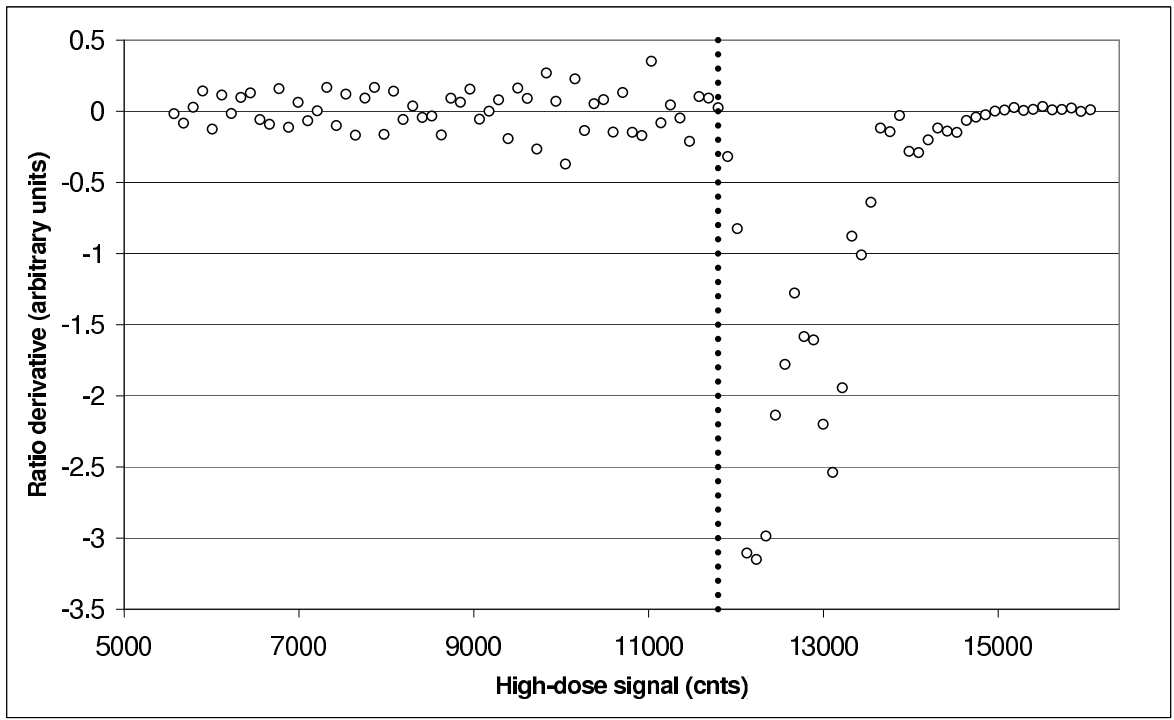}
%\vspace{-6cm}
\caption{\label{fig:Derivative}The derivative of the ratio of the pixel signals plotted against the high-dose signal. The 
vertical dotted line corresponds to the threshold value.}
\end{center}
\end{figure}

In order to extract the threshold from the input data, the following steps are taken.
\begin{enumerate}
\item The ratio (high-to-low-dose) of the pixel signals (corresponding to the same position on the detector) in the two 
images is histogrammed in bins of the high-dose signal; an average value is obtained in each bin. To safeguard against 
the introduction of noise, bins with less than ten contents are ignored. An example of one such distribution is shown in 
figure \ref{fig:RatioVSHighDose}. We observe that the ratio of the pixel signals is constant over a wide range of values 
of the high-dose signal, rapidly dropping at the place where signal saturation sets in.
\item The derivative of the ratio distribution is subsequently obtained (figure \ref{fig:Derivative}). The use of the 
derivative has advantages, namely, the riddance of possible slope effects in the original histogram and larger sensitivity 
in the area where the ratio of the pixel signals departs from constancy.
\item The threshold is obtained from the derivative plot as follows. Let us assume that an average of ratio-derivative 
values and an rms (standard deviation) have been calculated from a number of successive entries (bins); we will refer 
to these values as `current'. The value corresponding to the next bin in the plot (with increasing high-dose signal) 
is tested for constancy on the basis of the current average and rms; in case that the difference (of the bin value to 
the current average) is larger than fivefold the current rms (i.e., corresponding to a $5 \sigma$ effect for the normal 
distribution), the procedure terminates and the signal which is associated with the lower bound of the bin tested is 
originaly assigned to the threshold. The starting point in this iterative scheme is assumed to be the average and the rms 
calculated from the first ten entries; therefore, the eleventh bin is the first one to be tested for a possible deviation 
from constancy. The scanning of the derivative plot is done from left (low values of the high-dose signal) to right (high 
values of the high-dose signal).
\item The ratio-derivative plot is finally scanned backwards (decreasing high-dose signal) starting at the threshold 
determined in the previous step. A sign change in the difference of the bin value to the current average (compared 
to the sign of the corresponding difference in the case of the originally-assigned threshold) marks the position of the 
highest signal which still fulfils the linearity condition. This approach safeguards against several effects, including 
the existence of an occasional slope in the original plot (figure \ref{fig:RatioVSHighDose}), a drift in the average 
and rms values extracted on the basis of figure \ref{fig:Derivative}, etc. The method is robust; we are not aware of 
cases in which it failed to yield a reasonable output.
\end{enumerate}
In the implemented solution in the dual-dose imaging, the high-dose image is returned in case that no saturation could be 
detected (for example, as a result of the selection of inappropriate dose levels in the data acquisition).

\begin{figure}
\begin{center}
\includegraphics [height=15.5cm,angle=270] {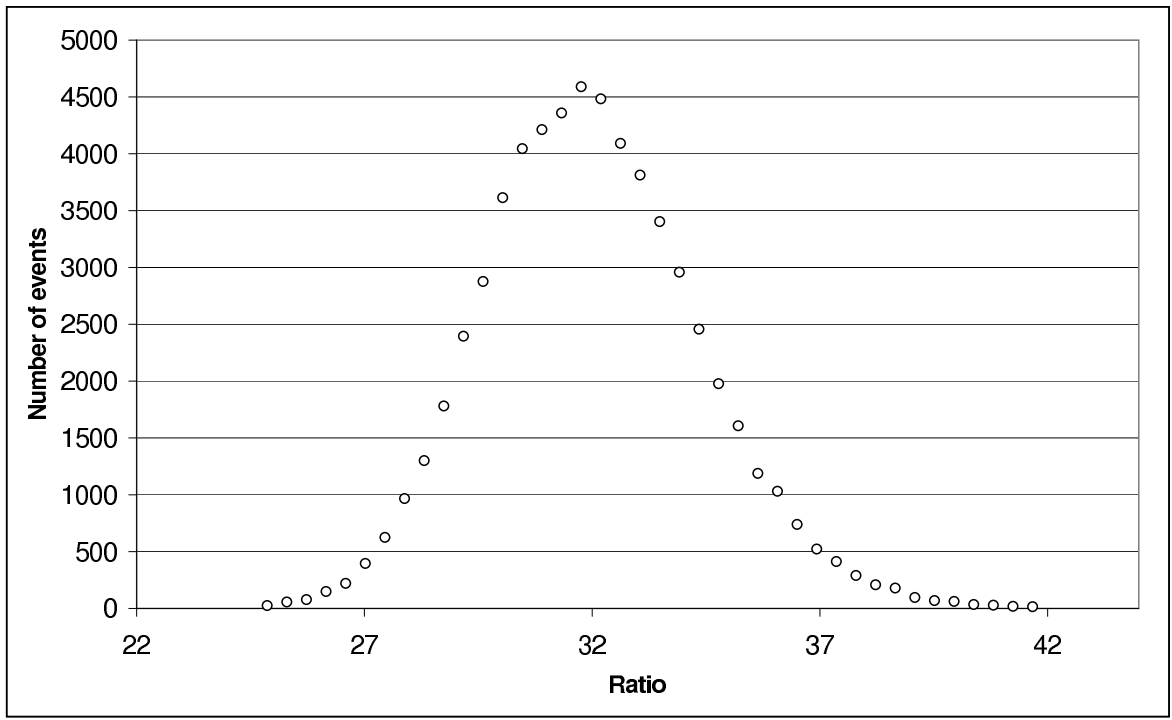}
%\vspace{-6cm}
\caption{\label{fig:Ratio}The distribution of the ratio of the pixel signals for all those pixels whose high-dose signal 
does not exceed the threshold value.}
\end{center}
\end{figure}

The optimal ratio of the pixel signals is subsequently obtained for all those pixels whose high-dose signal is below 
threshold (hence, unsaturated). Such a distribution is shown in figure \ref{fig:Ratio}; in case of less than ten entries, 
the bin values are neither shown nor taken into account in the statistics. The average of the distribution is obtained and 
used in the combination of the two images; the distribution of figure \ref{fig:Ratio} yields an average value of $32$ and 
an rms of $2.4$. In an ideal world, devoid of noise (quantum, readout, electronic, etc.), the ratio distribution should be 
a `$\delta$-function'.

\subsubsection{Combination of the two images.}

Having determined the threshold and ratio values, we can now proceed to the combination of the two images. Our strategy 
may easily be described in one sentence. If the pixel signal in the high-dose image is below the threshold, it is directly 
used; otherwise, it is substituted by the low-dose pixel signal after it has properly been scaled by the optimal ratio 
determined at the end of the previous section.
 
\begin{figure}
\begin{center}
\includegraphics [width=15.5cm] {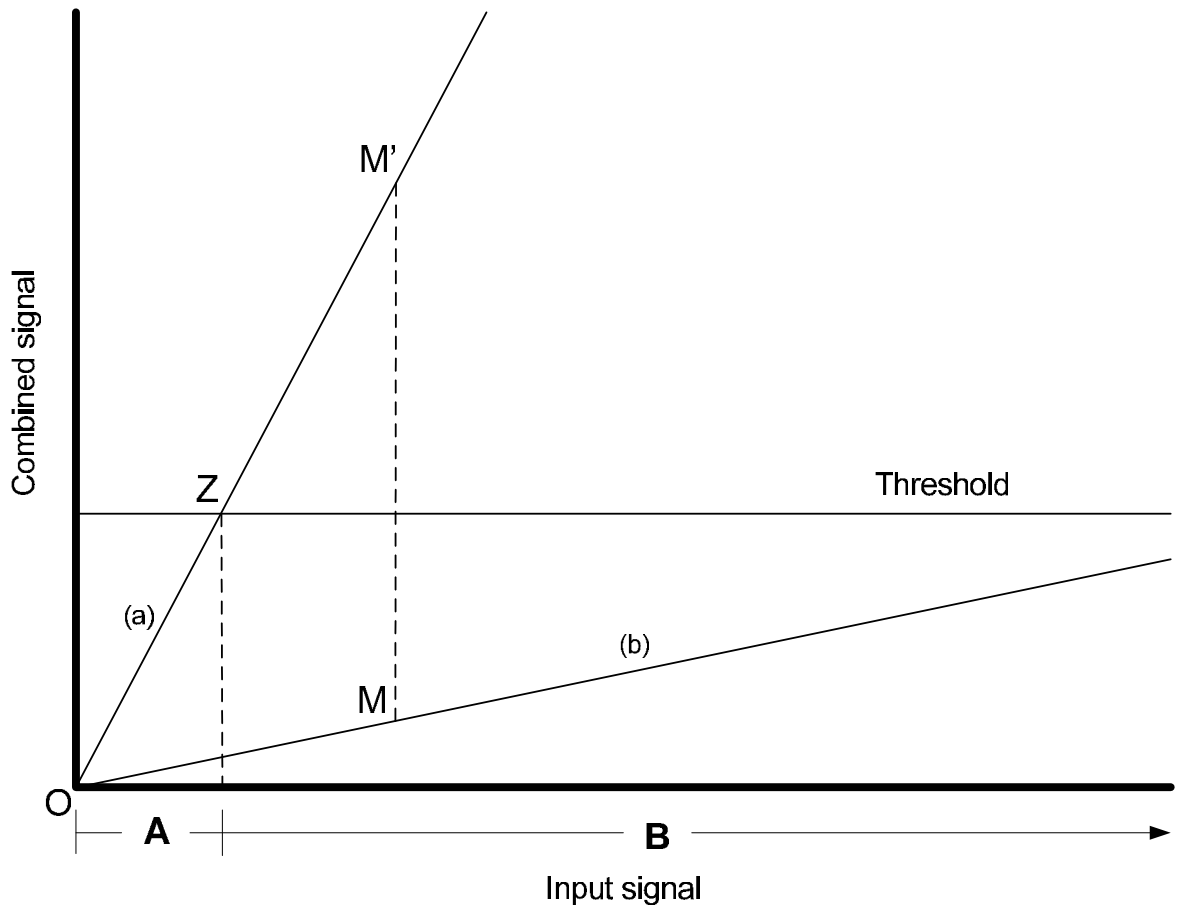}
%\vspace{-6cm}
\caption{\label{fig:DataProcessing}The strategy followed in the combination of the two images in schematic form. Up to 
the point where the high-dose signal (line (a)) saturates, it is used (OZ segment). Above the threshold value, the low-dose 
signal (line (b)) is used properly scaled; the scaling is done in such a way as to map point M onto M$^\prime$.}
\end{center}
\end{figure}

The combination of the two images may easily be understood with the help of figure \ref{fig:DataProcessing}. The high-dose 
signal (line (a)) is used in area A (where it is smaller than the threshold); the low-dose signal (line (b)) is used in 
area B (where the high-dose signal is useless). Finally, the combined image is constructed in the basis of the straight 
line OM$^\prime$; in the first segment (OZ), the high-dose signals are used, whereas in the second one (ZM$^\prime$), the 
low-dose signals are involved, along with the optimal ratio. Evidently, the dynamic range of the output signal (proportional 
to the length OM$^\prime$) is larger than the one corresponding only to the high-dose information (proportional to the length OZ).

\section{Results}

\begin{figure}
\begin{center}
\includegraphics [width=15.5cm] {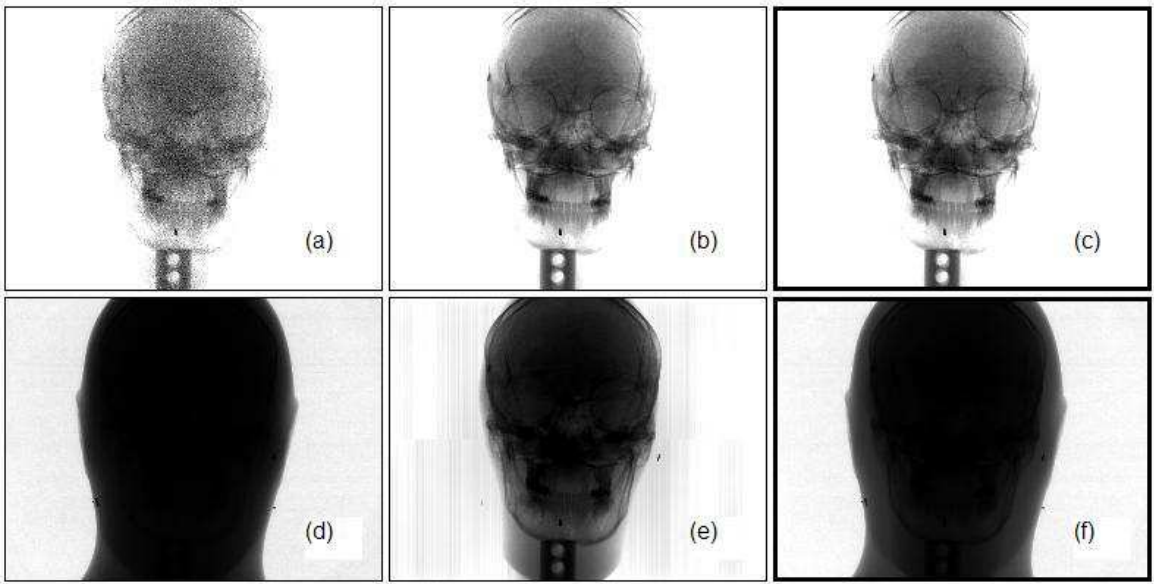}
%\vspace{-6cm}
\caption{\label{fig:HeadPhantom}The low-dose (a and d) and high-dose (b and e) images of the head phantom, along with the 
combined image (c and f). To show the usefulness of the approach proposed herein, the level and window values have been 
adjusted in such a way as to emphasise the details in the interior (a, b and c), as well as those close to the contour of 
the irradiated object (d, e and f). The degradation of the low-dose image is obvious in the interior of the image (a), 
while the high-dose image fails close to the contour of the irradiated object (e). The combined image reveals simultaneously 
the details in the interior (c) and around the contour of the irradiated object (f). The data in the inactive area of the 
detector are not shown in the case of the combined image (black borders).}
\end{center}
\end{figure}

The improvement in the image quality when using the dual-dose imaging technique is demonstrated in figure \ref{fig:HeadPhantom}. 
Two images of a head phantom were taken at $80$ kV and $25$ mA; the pulse-width settings were: $4$ msec for the low-dose image and 
$120$ msec for the high-dose image. Figures \ref{fig:RatioVSHighDose}-\ref{fig:Ratio} actually correspond to the analysis 
performed on these two images. The `theoretical' ratio of the pixel signals is expected to be equal to $30$ (ratio of the two 
pulse widths used), yet it came out closer to $32$ (figure \ref{fig:Ratio}) due to the fact that the acquisition pulse-width 
settings do not exactly match the properties of the actual pulses which are produced by the generator~\footnote{A correction 
scheme to account for this effect has been proposed and implemented in another problem, see Matsinos and Kaissl (2006).}.

It is evident from figure \ref{fig:HeadPhantom}a that the low-dose image is grainy; this is a good example of signal degredation 
as a result of quantum and discretisation noise. The high-dose image shown in figure \ref{fig:HeadPhantom}b does not suffer 
from this effect. On the other hand, the contour of the irradiated object is nicely reproduced in the low-dose image (figure 
\ref{fig:HeadPhantom}d); due to signal saturation, this is the place where the high-dose image fails (figure \ref{fig:HeadPhantom}e). 
The images shown in figures \ref{fig:HeadPhantom}c and \ref{fig:HeadPhantom}f correspond to the combination of the two images. 
The window and level parameters have been adjusted in figures \ref{fig:HeadPhantom} in such a way as to demonstrate the usefulness 
of the approach described in the present paper. In this example, the use of the two input images was quite balanced: $59.3 \%$ 
of the low-dose image is taken over in the combined image, $40.7 \%$ of the high-dose image is used. Evidently, details can 
be simultaneously seen both in the high- and low-attenuation areas of the irradiated object in the combined image.

\section{Conclusions}

The present paper introduces a method to increase the dynamic range of X-ray detectors which are used in imaging. The input data 
comprise two fully-corrected (for offset and gain effects) X-ray images of an object, obtained at different dose levels. In 
the current implementation, the intensity modulation is achieved through a variation of the width of the X-ray pulse; another 
option would be to vary the X-ray-tube current, or simultaneously both acquisition settings.

The data of the two input images are processed and finally yield the values of two parameters: the first parameter defines the 
level at which the switching from the high- to the low-dose information will occur in the combination of the two input images, 
whereas the second represents the amount by which the low-dose image will be scaled whenever it is used. A robust method is 
proposed to extract reliably the values of these two parameters from the input images. The novelty of the approach relates to 
the fact that the parameters, needed in the combination of the input images, are determined exclusively from the information 
these images contain; therefore, no additional calibrations (save for the standard ones, leading to the offset and gain 
corrections) are needed.

Finally, one image is created, containing signals which are selectively obtained from either of the input images; the dynamic 
range of the output image is larger than the one corresponding to the input images (which is characterised by the limitation 
in the discretisation depth of the detector). The final image contains useful information in all its entirety. The selective 
use of parts of the two input images allows both the contour of the irradiated object to be visible and the high-attenuation 
areas to retain their image quality corresponding to the high-dose image. The benefits of the method have been convincingly 
demonstrated with an example involving a head phantom.

\begin{ack}
The data analysed in this paper were acquired by H Riem.
\end{ack}

\References
\item[] Dasani G, Kaissl W, Matsinos E, Morf D and Riem H 2006 Dual pulse imaging {\it U.S. Patent Application} 
{\bf 20060724667}
\item[] Eberhard J W, Claus B E H and Landberg C 2005 Enhanced X-ray imaging system and method {\it U.S. Patent 
Application} {\bf 20050226375}
\item[] Jabri K N, Avinash G B, Rader A E, Uppaluri R, Sabol J M and Nicolas F S 2004 Method, system and computer 
product for processing dual energy images {\it U.S. Patent} {\bf 6816572}
\item[] Kump K S and Odogba J 2003 Method and system for dual or multiple energy imaging {\it U.S. Patent Application} 
{\bf 20030169850}
\item[] Matsinos E and Kaissl W 2006 The dual-gain mode: a way to enhance the dynamic range of X-ray detectors {\it Preprint} 
physics/0607021
\item[] Zhao J, HibbsOpsahl-Ong B and Hopple M R 2004 Dual energy x-ray imaging system and method for radiography and 
mammography {\it U.S. Patent} {\bf 6683934}
\endrefs

\clearpage
% ============= FIGURE1

\clearpage
% ============= FIGURE2

\clearpage
% ============= FIGURE3

\clearpage
% ============= FIGURE4

\clearpage
% ============= FIGURE5

\end{document}